# Microwave magnetic excitations in U-type hexaferrite Sr$_4$CoZnFe$_{36}$O$_{60}$ ceramics

M. Kempa,[1a)] V. Bovtun,[1] D. Repček,[1,2] J. Buršík,[3] C. Kadlec,[1] S Kamba[1]

[1]Institute of Physics of the Czech Academy of Sciences, Na Slovance 2, Prague 182 00, Czech Republic

[2]Faculty of Nuclear Sciences and Physical Engineering, Czech Technical University in Prague, Břehová 7, 115 19 Prague, Czech Republic

[3]Institute of Inorganic Chemistry, Czech Academy of Sciences, 250 68 Řež near Prague, Czech Republic

**Abstract.**

Microwave (MW) transmission, absorption, and reflection loss spectra of the ferrimagnetic U-type hexaferrite Sr$_4$CoZnFe$_{36}$O$_{60}$ ceramics were studied from 100 MHz to 35 GHz at temperatures between 10 and 390 K. 9 MW magnetic excitations with anomalous behavior near the ferrimagnetic phase transitions were revealed. They also change under the application of weak bias magnetic field (0 – 700 Oe) at room temperature. 6 pure magnetic modes are assigned to dynamics of the magnetic domain walls and inhomogeneous magnetic structure of the ceramics, to the natural ferromagnetic resonance (FMR) and to the higher-frequency magnons. Three modes are considered as the magnetodielectric ones with dominating influence of the magnetic properties on their temperature and field dependences. Presence of the natural FMR in all ferrimagnetic phases proves existence of the non-zero internal magnetization and magnetocrystalline anisotropy. Splitting of the FMR into the two components without magnetic bias was observed in the collinear phase and is attributed to a change of the magnetocrystalline anisotropy during the phase transition. The high-frequency FMR component critically slows down to the phase transition. At room temperature, the FMR splitting and essential suppression of the higher-frequency modes was revealed under the weak bias field (300 – 700 Oe). The highly nonlinear MW response and the FMR splitting are caused by the gradual evolution of the polydomain magnetic structure to a monodomain one. The high number of magnetic excitations observed in the MW region confirms the suitability of using hexaferrite Sr$_4$CoZnFe$_{36}$O$_{60}$ ceramics as MW absorbers, shielding materials and highly tunable filters.

a) Author to whom correspondence should be addressed: kempa@fzu.cz

# 1. Introduction

Ferrite materials, called hexaferrites, exhibit a complicated ferrimagnetic arrangement with a hexagonal crystal structure that can be described as stacking sequences of three basic blocks: S $Me_2Fe_4O_8$, (spinel block), where $Me$ denotes divalent metal ion; R $[(Ba,Sr)Fe_6O_{11}]^{2-}$, (hexagonal block) and T $(Ba,Sr)_2Fe_8O_{14}$, (hexagonal block). Depending on the stacking sequences of these blocks, hexaferrites are divided into six main types [1,2]. Typical examples are M-type hexaferrite $BaFe_{12}O_{19}$ (stacking sequence SR), Y-type $BaSrCoZnFe_{11}AlO_{22}$ (ST), W-type $SrZnCoFe_{16}O_{27}$ (SSR), Z-type $(Ba,Sr)_3Co_2Fe_{24}O_{41}$ (STSR), X-type $Ba_2Co_2Fe_{28}O_{46}$ (SRS*S*R) and U-type $Sr_4Co_2Fe_{36}O_{60}$ (SRS*R*S*T), where the (*) symbol means that the corresponding block has been turned 180° around the hexagonal c-axis. Due to their unique magnetic properties and economic availability, hexaferrite materials (M-hexaferrites in particular) are often used in engineering practice, e.g., in refrigerator seal gaskets, microphones, loudspeakers, small motors, clocks, magnetic separators and in microwave generators, filters and absorbers [3,4].

Hexaferrites are also very interesting from the scientific point of view. Nearly twenty years ago, Kimura found that the conical magnetic structure in Y-hexaferrite induces a ferroelectric polarization, and that such a multiferroic system exhibits strong magnetoelectric coupling [5]. Similar spin-induced ferroelectricity has been uncovered in a variety of materials with different hexaferrite crystal structures, which often exhibit strong magnetoelectric coupling not only at helium temperatures, but also at and above room temperature (RT) [6,7,8]. Depending on the chemical composition, these materials are ferrimagnetic up to relatively high temperatures of 600-750 K and generally exhibit a set of magnetic phase transitions upon cooling from collinear structures to various conical ones, among which the transverse and alternating longitudinal conical structures are multiferroic [1]. Since hexaferrites exhibit spin-induced ferroelectricity, their magnetoelectric coupling in small magnetic fields is several orders of magnitude higher than in classical $BiFeO_3$, which belongs to type-I multiferroics, where linear magnetoelectric coupling is forbidden due to the incommensurately modulated cycloidal magnetic structure. Only in the field above 18 T does the magnetoelectric coupling in $BiFeO_3$ increase dramatically due to the transition to the canted antiferromagnetic state, where linear magnetoelectricity is already allowed [9]. Static magnetoelectric coupling has so far been studied mainly in hexaferrites with Y- and Z-crystal structures, but multiferroicity has been reported also in some hexaferrites with M-, W-and U-hexaferrite structures [10,11,12,13]. Interestingly, despite the chemical and structural affinities of different hexaferrites, the magnetoelectric coupling in Y- and Z-hexaferrite arises from different origins: inverse Dzyaloshinskii-Moriya interaction ($\propto \mathbf{S}_i \times \mathbf{S}_j$) is responsible for the static magnetoelectric coupling in the Y-type hexaferrite, while additional *p-d* hybridization [$\propto (\mathbf{e}_i \cdot \mathbf{S}_j)^2 \mathbf{e}_i$; here $\mathbf{S}_i$ and $\mathbf{e}_i$ denote *i*-th spin and the bond direction, respectively] becomes dominant in the Z-type hexaferrite [14].

A dynamic magnetoelectric coupling can be responsible for the activation of spin excitations in microwave (MW) and terahertz (THz) dielectric spectra. Such excitations are called electromagnons and in hexaferrites they are activated, regardless of the type of static magnetoelectric coupling, by exchange striction ($\propto \mathbf{S}_i \cdot \mathbf{S}_j$) [15,16,17,18,19,20]. One of the characteristic features of an electromagnon is a directional dichroism: it means that the reversal of the light incidence direction gives a different electromagnon absorption [21]. Such an effect has mainly been observed in the THz region but has recently been discovered also in the MW region for Y hexaferrite [22]. In this case the sign of nonreciprocal MW response could be controlled by the poling electric field, which opens a new avenue for practical applications in magnonics and other future wireless communication technologies. This motivated us to study the microwave properties of $Sr_4CoZnFe_{36}O_{60}$ (CoZnU) with U-type hexaferrite structure.

The microwave dielectric permittivity and magnetic permeability of various materials with U-hexaferrite structure have been published more frequently, but practically only at RT and in a limited frequency range up to 6, 12 or maximum 18 GHz [23,24,25,26]. In our paper, we focus on CoZnU, which is chemically and structurally related to multiferroic $Sr_4Co_2Fe_{36}O_{60}$ (trigonal space group $R\bar{3}m$) [27]. Very recently, we reported that it exhibits a transition from a paramagnetic to a collinear ferrimagnetic structure at $T_{c1}$ = 635 K [28]. At $T_{c2}$ = 305 K, it transforms into a first conical magnetic structure and at $T_{c3}$ = 145 K most probably into a second one, yet not clearly identified [28]. Dielectric studies did not reveal any anomalies near magnetic phase transitions and magnetoelectric measurements did not find any intrinsic ferroelectric polarization at 10 K [28]. It means that CoZnU is not multiferroic. Here we report the results of our MW investigations performed up to 50 GHz and down to 10 K, which reveal 9 magnetic excitations, including a ferromagnetic resonance. Most of these excitations show strong temperature dependences near the magnetic phase transitions and we will discuss their origin.

## 2. Experimental details

For synthesis of the CoZnU powders and ceramics, the Pechini-type in situ polymerizable complex method relying on the polycondensation reaction (polyesterification) between citric acid and ethylene glycol was used. Calculated amounts of the strontium carbonate ($SrCO_3$), cobalt nitrate ($Co(NO_3)_2·6H_2O$), zinc nitrate ($Zn(NO_3)_2·6H_2O$), and iron nitrate ($Fe(NO_3)_3·9H_2O$; all chemicals from Sigma-Aldrich, 99.9 % purity) were dissolved in the distilled water and reacted at 403 K with a solution of a polymer gel formed by the reaction between citric acid ($HOOCCH_2C(OH)$-$(COOH)CH_2COOH$) and ethylene glycol ($HOCH_2CH_2OH$) in water leading to the voluminous polymer resin. After the resin breaking, drying (at 423 K) and charring (at 623 K) it for 24 h, the powder was heat-treated in an oxygen atmosphere at 1488 K for 12 h. A cold isostatic pressing (300 MPa/300 s) and subsequent sintering at 1488 K in oxygen atmosphere led to dense ceramics of the U-phase hexaferrite. X-ray diffraction revealed 97 % pure U-hexaferrite with $R\bar{3}m$ space group symmetry and lattice parameters ($a$ = 5.871 Å, $c$ = 112.594 Å) consistent with the parameters of $Sr_4Co_2Fe_{36}O_{60}$. A detailed description of the CoZnU ceramics preparation, structure and characterization of its static and dynamic dielectric and magnetic properties from can be found in ref. [28].

In this report, the MW spectroscopy study was performed using two kinds of planar lines: a microstrip line (MSL) and a grounded coplanar waveguide (CPW) with different electromagnetic field distribution and coupling between the sample and the line. Our choice is caused by the following reasons. The precise and highly accurate CPW and MSL test boards, which are commercially available and provide propagation of the only quasi-TEM wave without any resonance reflections below 50 GHz [29], were used. Our study is focused on the absorption bands characterizing the measured material. The frequency range (1 – 50 GHz), 50 Ohm impedance, geometry and size of the MSL and CPW test boards correspond well to our experimental facilities and sample size. The propagating quasi-TEM wave provides a good and clear coupling of the electromagnetic field with a sample.

The full set of spectra of the scattering $S$-parameters ($S_{11}$, $S_{22}$, $S_{21}$, $S_{12}$, both magnitude and phase) [30] of the Southwest Microwave MSL and CPW 50 GHz test boards [31] loaded with the parallel plate samples (~ 1 mm thick, ~ 100 mm$^2$ surface) at their top (Fig. 1) were recorded in the transmission setup between 100 MHz and 50 GHz on heating from 10 K to 390 K in a Janis closed-cycle He cryostat with a temperature rate of 0.5 K/min using the Agilent E8364B network analyzer. The influence of the bias magnetic field $H$ (up to 700 Oe) on the $S$- parameters spectra, reflection loss, magnetic nonlinearity and magneto-impedance effect were studied at RT in the MSL and CPW setups. Reflection loss ($S_{11}$ parameter in the reflection setup [30]) spectra were measured between 100 MHz and 20 GHz using the "short" calibration standard.

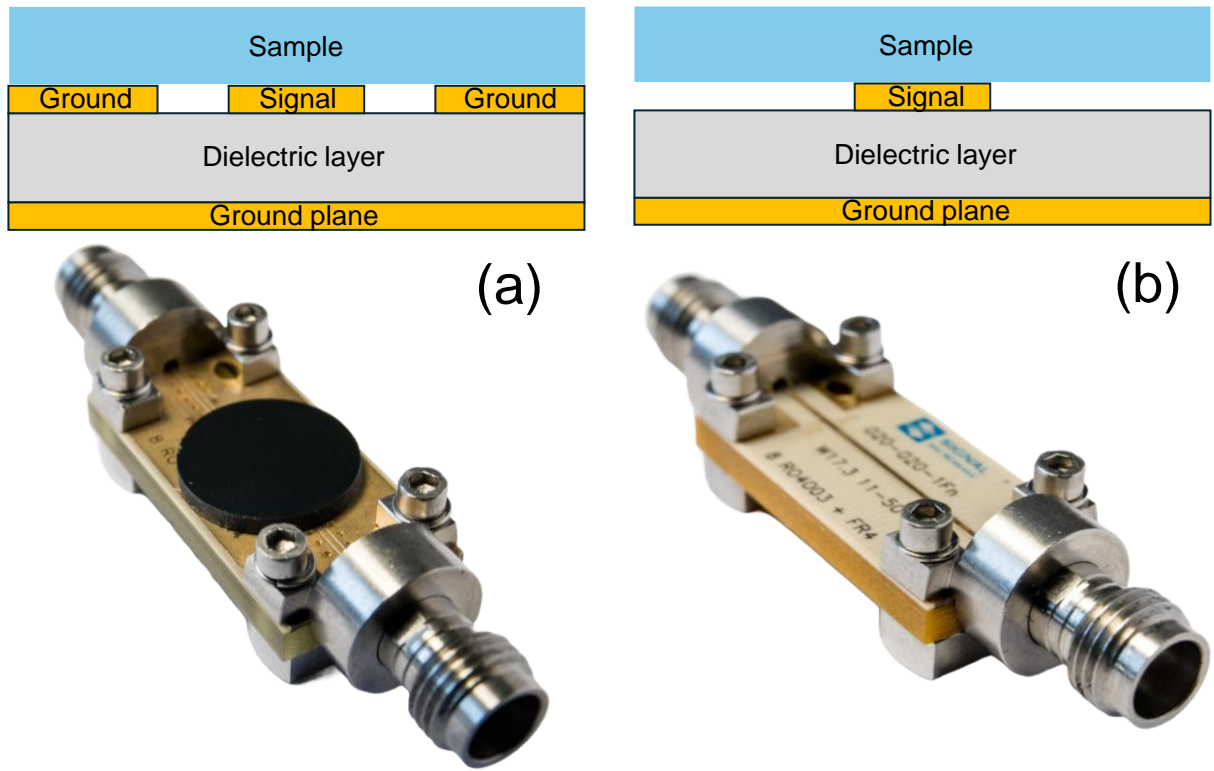

Fig. 1. (a) Schematical cross-section view and image of the CPW test board with a sample; (b) schematical cross-section view of the MSL test board with a sample and its image without a sample. Samples, signal electrodes and ground electrodes are shown in the cross-section view with different colors. The quasi-TEM wave propagates along the test board, both electrical and magnetic components of the MW field are in the cross-section plane.

The resonance MW measurements were performed using the thin circular disk sample (~0.22 mm thick, ~12 mm in diameter) as a dielectric resonator (DR) with a few resonance frequencies [32] between 8 and 16 GHz, and as a part of the composite dielectric resonator (CDR) [32,33,34] at the $TE_{01\delta}$ resonance near 5.8 GHz. The resonators were measured in a cylindrical shielding cavity using the transmission setup with a weak coupling by an Agilent E8364B network analyzer during heating from 10 to 390 K with a temperature rate of 0.5 K/min in a Janis closed cycle He cryostat. In the first case, $S_{21}$ (transmission coefficient) was recorded. In the latter case, the $TE_{01\delta}$ resonance frequency, quality factor, and insertion loss of the base cylindrical dielectric resonator with and without sample were recorded. The effective refractive index and loss tangent of the samples were calculated from the measured parameters of the CDR using the MATLAB-based software [32], accounting for volume fractions of the CDR parts [33,34].

Time-domain THz spectroscopy was performed by measuring the complex sample transmittance using a custom-made spectrometer based on a Ti:sapphire femtosecond laser oscillator (Coherent, Mira, 800 nm, 80 fs, 8 nJ, 76 MHz). Linearly polarized THz pulses were generated by an interdigitated photoconducting antenna (TeraSED, GigaOptics). They were detected by the method of the electro-optic sampling in a 1-mm thick ⟨110⟩-oriented ZnTe crystal. The accessible spectral range of the spectrometer is 0.2 to 3 THz, but our spectra were obtained only up to 1.7 THz because the sample becomes opaque at higher frequencies due to strong phonon absorption. The low-temperature measurements were realized in the transmission geometry under normal incidence in a helium-flow cryostat (Optistat, Oxford Instruments) with mylar windows.
Knowing the thickness of the measured ceramics (0.22 mm), we calculated its complex refractive index. In the considered interval, the uncertainty was mainly linked to the precision in the measurement of the sample thickness and remained below 3 %. The Gouy shift correction was taken into account [35].

## 3. Broadband microwave transmission and absorption.

The recorded full set of *S*-parameters of the MSL and CPW loaded with CoZnU ceramics was analyzed (see Fig. S1 and details in Supplementary material [36]). As the most sensitive and informative, the following parameters were selected and calculated [30,37,38]: $S_{21}$ magnitude in dB (transmission coefficient *Tr*), input impedance $Z_{in}$ and absorption coefficient *A*. *Tr* is the most reliable directly measured *S*-parameter. Its recorded spectra with multiple resonances are similar for the CPW and MSL experiments and show qualitatively similar temperature changes. This similarity is well demonstrated by the comparison of the *Tr(f,T)* temperature-frequency maps (Fig. 2 and Fig. S2 in Supplementary material). In more detail, the temperature evolution of broadband *Tr(f)* spectra is presented in Fig. 3 for the CPW, and in Fig. S3 of Supplementary material [36] for the MSL.

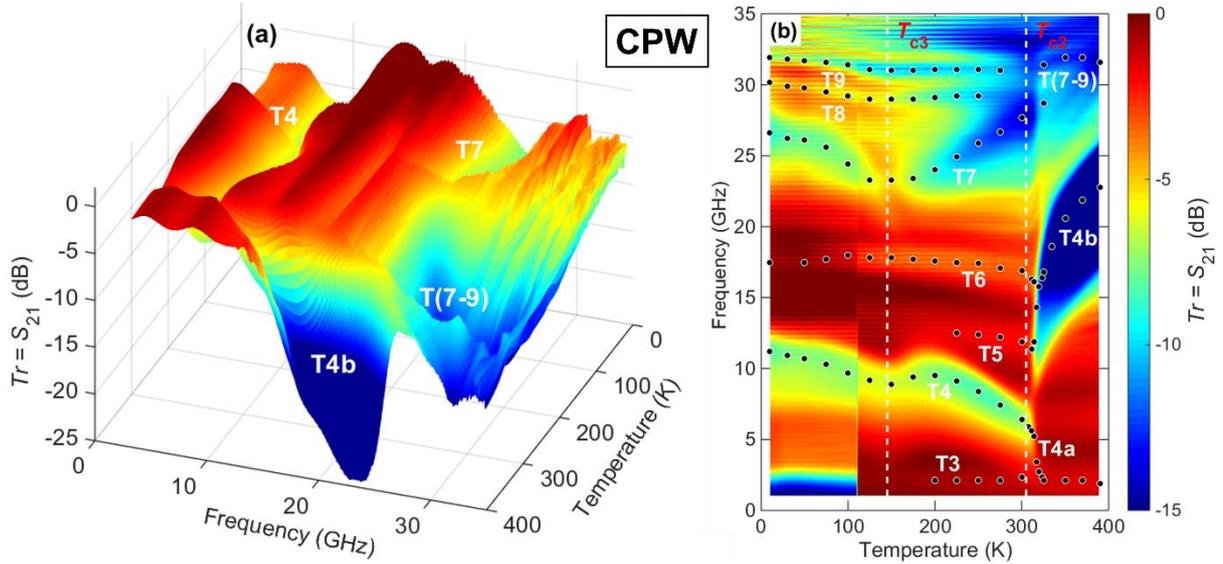

Fig. 2. Temperature-frequency maps of the transmission coefficient $Tr$ ($S_{21}$ magnitude) of CoZnU ceramics obtained using the CPW: (a) in 3D view, (b) as a pseudo-3D color map. Similar maps acquired with the MSL are shown in Fig. S2 of the Supplementary material [36]. In (b), points correspond to the main $Tr(f)$ minima shown in Fig. 3. Temperatures of the phase transitions are marked by dashed lines. Apparent jumps of $Tr$ at several temperatures are instrumental artifacts.

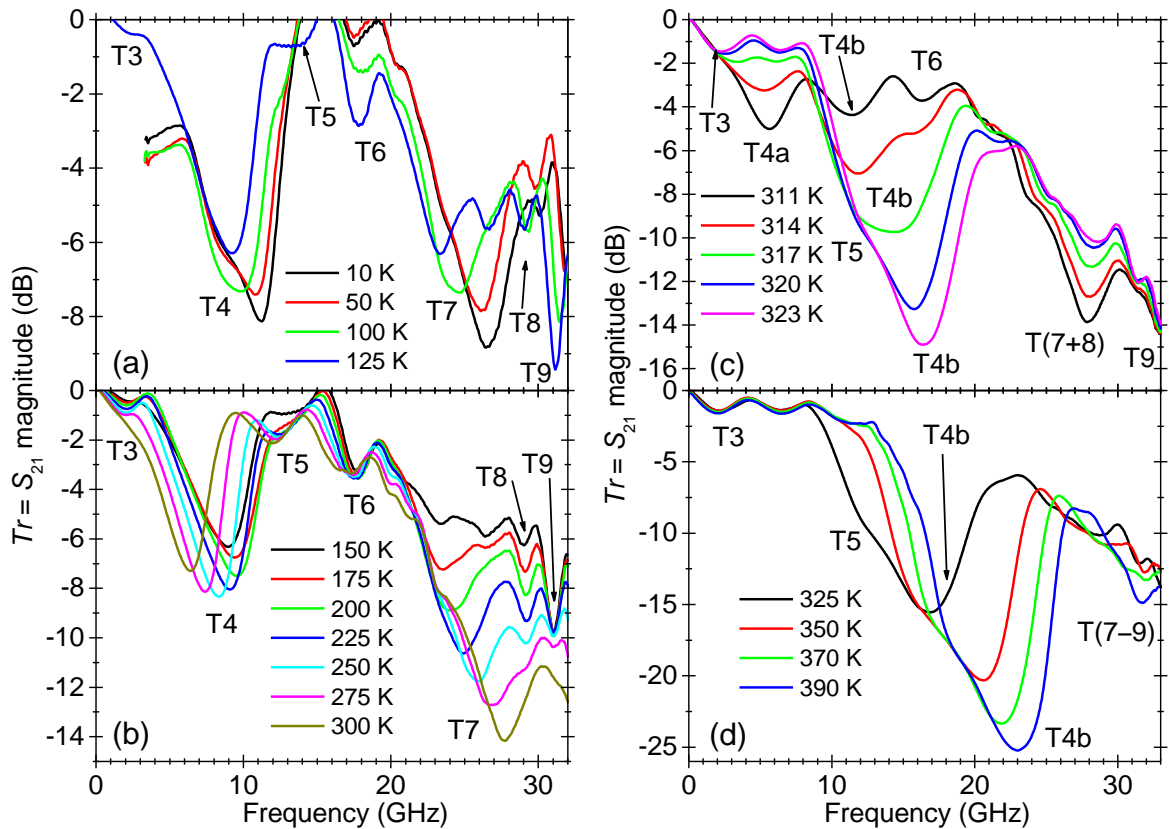

Fig. 3. Broadband spectra of the transmission coefficient $Tr$ ($S_{21}$ magnitude) of the CPW loaded with a CoZnU sample at selected temperatures below $T_{c3}$ (a), between $T_{c3}$ and $T_{c2}$ (b), and above $T_{c2}$ (c, d). The main $Tr(f)$ minima are marked T3–T9.

We limited our analysis to the frequency range 1–35 GHz because of a too high density of the observed resonances and noise level of the $Tr(f)$ spectra above 35 GHz (see Fig. S1 in Supplementary material [36]). Below 35 GHz, 7 well-defined resonance-like or diffuse $Tr(f)$ minima were selected and numbered from T3 to T9 with increasing frequency (here we continue in the numbering of the excitations seen below 2 GHz in the complex magnetic permeability $\mu^* = \mu' + i\mu''$ published in [28]). Accounting for the observed splitting of the T4 minimum above $T_{c2}$, its two components are labeled as T4a and T4b. The lower-frequency T4a mode joins the T3 one on further heating (above 320 K), the higher-frequency F4b mode critically slows down towards $T_{c2}$ on cooling. All mentioned $Tr(f)$ minima correspond to the MW excitations (transmission modes, T-modes) of the MW line loaded with a CoZnU sample. The same set of T-modes is observed in both the CPW and MSL measurements of the same CoZnU sample. Moreover, the temperature evolution of the CPW and MSL T-modes is similar, with a remarkable change of their behavior near the CoZnU magnetic phase transitions. The T7, T8 and T9 modes are well separated below $T_{c2}$, but become overlapped above $T_{c2}$, forming one T(7-9) mode. The T4b mode interferences with the T5 and T6 ones above $T_{c2}$ (between 305 K and 320 K). As a result, corresponding $Tr(f)$ minima join into the diffuse and asymmetric one, T5 and T6 gradually disappear from the transmission spectra. Consequently, identification of the T3, T4a, T4b, T5 and T6 transmission modes in the collinear phase close to the phase transition is complicated. Nevertheless, we carefully followed their temperature evolution: we acquired the transmission spectra with the 1 K step during slow heating (0.5 K per minute). So, we are generally sure about our mode attribution. The T4b mode dominates above $T_{c2}$. Two modes, T4 (T4a, T4b above $T_{c2}$) and T7, are characterized by a very pronounced temperature dependence.

The analysis of the transmission coefficient spectra $Tr(f)$ allowed us to observe, reliably separate and study several MW excitations in the CoZnU hexaferrite. For a better understanding of their nature, the analysis of the absorption coefficient spectra $A(f)$ could be very useful. The $A(f)$ spectrum characterizes the absorption of the electromagnetic energy in a sample. Consequently, $A(f)$ maxima characterize the frequency ranges of the increased absorption of electromagnetic energy due to the interaction with the excitations in the studied material. While the transmission coefficient $Tr$ characterizes the electromagnetic energy which was not absorbed in a sample and not reflected from the sample, the frequencies of the $Tr(f)$ minima are defined by the superposition of the reflection and absorption maxima. Only in the case where the absorption prevails, the frequencies of the $Tr(f)$ minima correspond to those of the $A(f)$ maxima, both being related to frequencies of the material loss maxima and could be reasonably used for the physical interpretation of the MW excitation, for the study of their microscopic origin. The calculated (see Supplementary material [36]) $A(f)$ spectra are shown for the CPW experiment in Fig. 4 and for the MSL experiment in Fig. S4 in Supplementary material [36].

Let us note that the deep and broad transmission minima and high-level and broad absorption maxima observed in the MW region evidence the suitability of using the CoZnU hexaferrite ceramics as the microwave absorbing and shielding materials [3,4].

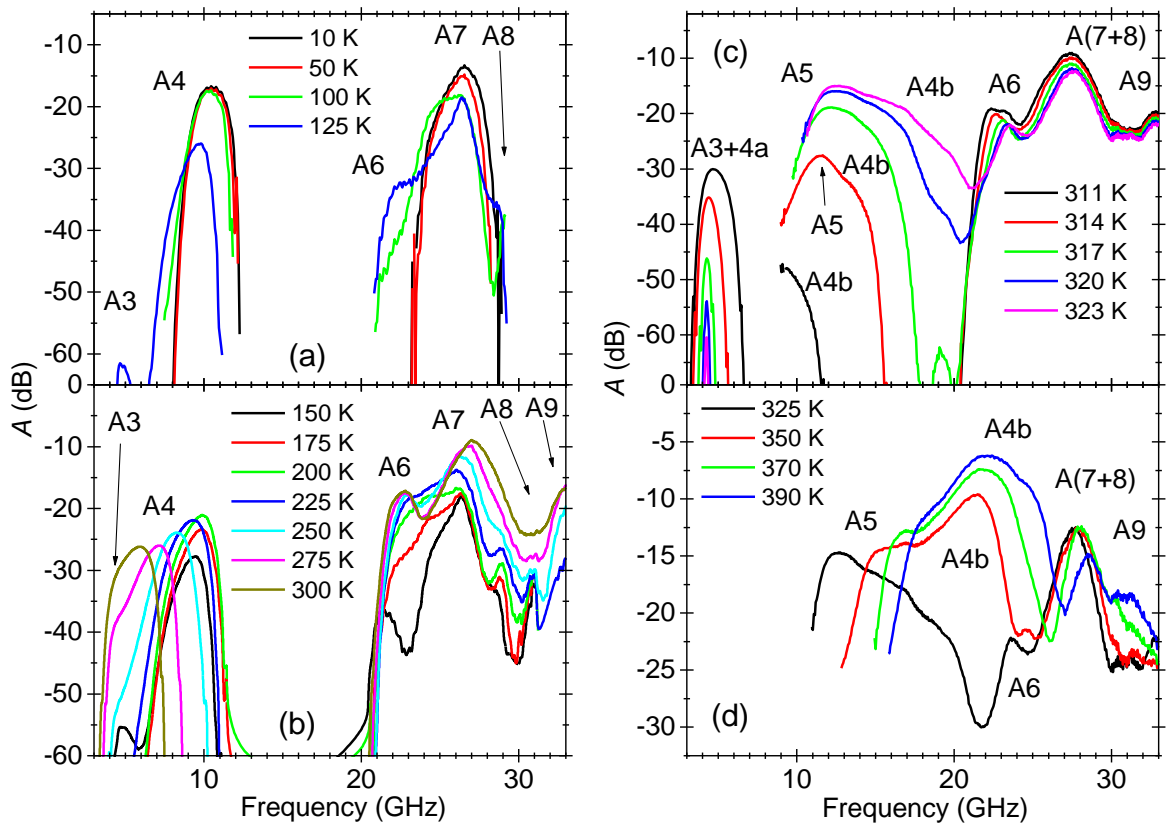

Fig. 4. Broadband spectra of the absorption coefficient $A$ of the CPW loaded with a CoZnU ceramics at selected temperatures below $T_{c3}$ (a), between $T_{c3}$ and $T_{c2}$ (b), and above $T_{c2}$ (c, d). The main $A(f)$ maxima are marked A3–A9.

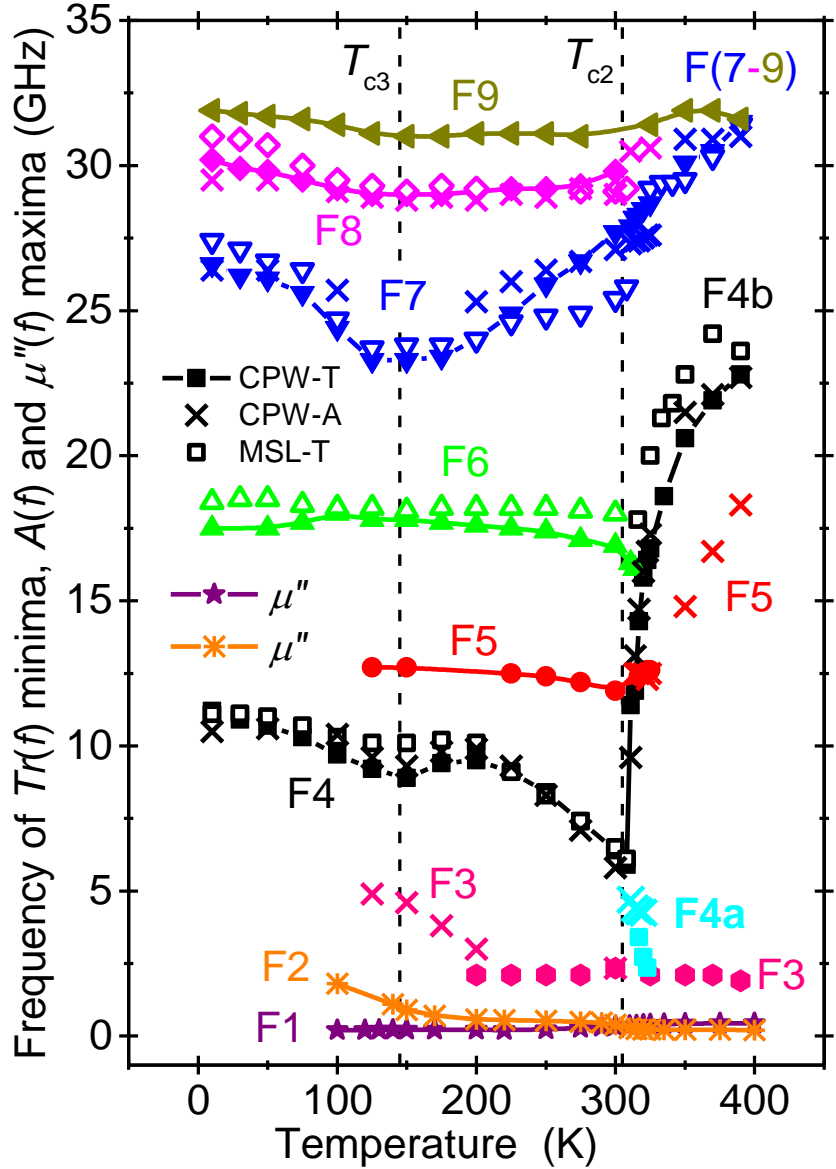

Fig. 5. Temperature evolution of the mean frequencies of the MW excitations F3–F9 corresponding to $Tr(f)$ minima and $A(f)$ maxima of the CPW (lines with filled symbols and crosses, respectively) and MSL (empty symbols) loaded with a CoZnU sample. The mean frequencies of the magnetic excitations F1 and F2, corresponding to the $\mu''(f)$ maxima from coaxial measurements below 2 GHz and published in [28], are also added.

The absorption spectra are noisier than the transmission ones, nevertheless 7 diffuse $A(f)$ maxima can be recognized below 35 GHz and numbered from A3 to A9 with increasing frequency. These maxima evidence 7 MW excitations (absorption modes, A-modes) of the CoZnU hexaferrite. The same set of the A-modes and their similar temperature evolution were derived from both CPW and MSL experiments. The A-mode system corresponds to the T-mode one. The frequencies of the A-modes remarkably change near the magnetic phase transitions, like those of the T-modes. Therefore, we consider all observed modes (T3–T9, A3–A9) to be related to the magnetic properties and MW excitations (F3–F9) of the CoZnU hexaferrite. We also analyzed the input impedance spectra $Z_{in}(f)$, see Fig. S5 in Supplementary material [36]. The main $Z_{in}(f)$ maxima also correspond to the frequencies of the F3–F9 excitations. The temperature evolution of the F3–F9 excitations is presented in Fig. 5 together with the F1 and F2 magnetic excitations, corresponding to the $\mu''(f)$ maxima from coaxial measurements below 2 GHz and published in [28].

The transmission spectra of the thin circular disk of the CoZnU sample, measured as a dielectric resonator (DR) in a shielding cavity [30,32], show a few temperature-dependent resonances with an anomalous behavior near $T_{c2}$ = 305 K (see temperature-frequency maps in Fig. S5 of Supplementary material [36]). This confirms critical dynamics of the MW excitations in the CoZnU ceramics. We should note that the DR resonance modes are magnetodielectric ones. Their resonance frequencies are defined by the refractive index, i.e. by both the dielectric permittivity and magnetic permeability of the material. As far as the dielectric permittivity is only slightly temperature dependent above 1 GHz, without any anomalies at the magnetic phase transitions [28], temperature dependences of the DR modes, with anomalies near $T_{c2}$, are related to the critical dependence of the MW magnetic permeability (at least, up to 13 GHz). See also the next section.

An essential fact is that independently of the used measurement setup (CPW, MSL or DR) and of the type of coupling, the main resonance modes in transmission, absorption or impedance MW spectra exhibit a similar temperature behavior with anomalies near the phase transitions. This shows that the MW excitations seen in Fig. 5 are indeed genuine in the CoZnU ceramics, and we will explain their origin in the next section.

**4. Temperature evolution of the microwave magnetic excitations.**

No anomalies of the complex dielectric permittivity $\varepsilon^* = \varepsilon' - i\varepsilon''$ (measured between 1 Hz and 1 GHz) were observed near the ferrimagnetic phase transitions at $T_{c3}$ = 145 K and $T_{c2}$ = 305 K [28]. It is true that at low frequencies a colossal permittivity was observed in CoZnU [28], which could mask the dielectric anomaly at the spin-induced ferroelectric phase transition, but this colossal permittivity is due to the Maxwell-Wagner relaxations from inhomogeneous conductivity in the CoZnU ceramics. Therefore, the permittivity decreases with increasing frequency, and at 100 MHz or 1 GHz it already takes intrinsic values of 20-30 (see Fig. 6a). The absence of the dielectric anomalies at $T_{c2}$ and $T_{c3}$, together with the fact that no intrinsic ferroelectric polarization was observed [28], suggest that CoZnU is not a multiferroic at all temperatures below $T_{c1}$.

On the other hand, temperature dependences of the permeability $\mu'(T)$ and loss $\mu''(T)$ at 100 MHz and 1 GHz are characterized by the remarkable maxima at $T_{c3}$ and $T_{c2}$. Below 3 GHz, we separately measured the permittivity and the permeability using coaxial techniques [28]. At higher frequencies, the dielectric and magnetic parameters cannot simply be distinguished [32]. Our MW measurements, using the MSL, CPW, DR and CDR techniques, as well as the THz one, characterize the full response of the studied material on both electric and magnetic components of the applied electromagnetic field, which depends on both material's permittivity and permeability. Our MSL and CPW techniques revealed the temperature evolution of the main MW excitation in the material, but they do not allow the quantitative estimation of the dielectric and magnetic parameters. The experimental data of the CDR and THz techniques can be analyzed [33,34] using the following mixed electromagnetic parameters: the product $\varepsilon'\mu'$ having the meaning of the square of the refractive index (for negligible index of absorption), and $\varepsilon''\mu''/\varepsilon'\mu'$ as a parameter characterizing electromagnetic losses (loss tangent) in the material. [39]. We also calculated $\varepsilon'\mu'$ and $\varepsilon''\mu''/\varepsilon'\mu'$ parameters for 1 GHz (using permittivity and permeability values from coaxial measurements [28]) and show all of them in Fig. 6b,c.

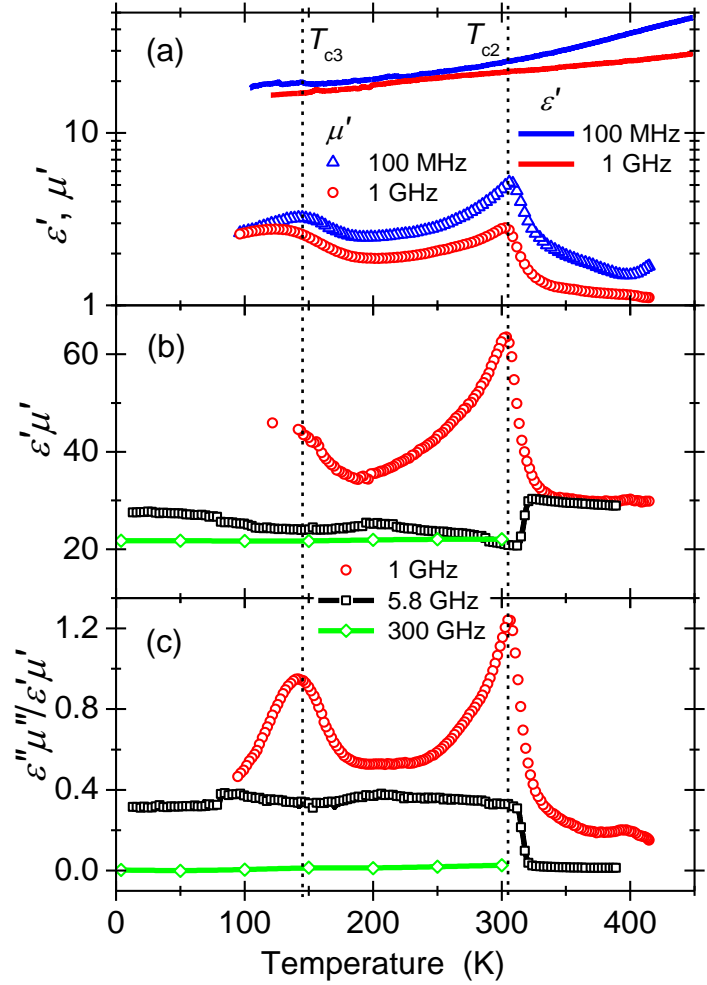

Fig. 6. Temperature dependences of the (a) $\varepsilon'$ (lines) and permeability $\mu'$ (symbols) at 100 MHz and 1 GHz obtained by coaxial techniques and published in [28]; (b) and (c) $\varepsilon'\mu'$ and $\varepsilon''\mu''/\varepsilon'\mu'$ electromagnetic parameters at 1, 5.8 and 300 GHz, respectively.

The comparison of the 1 GHz $\varepsilon'\mu'$ parameter (Fig. 6b) with $\varepsilon'$ and $\mu'$ (Fig. 6a) proves that the $\varepsilon'\mu'$ maxima at $T_{c3}$ = 145 K and $T_{c2}$ = 305 K are caused by the $\mu'(T)$ anomalies. The absence of the dielectric loss $\varepsilon''$ maxima in the 1 GHz temperature dependence [28] evidences an exclusive magnetic nature of the $\varepsilon''\mu''/\varepsilon'\mu'$ loss maxima in the 1 GHz curve (Fig. 6c). Since the 300 GHz $\varepsilon'\mu'$ value is nearly equal to the 1 GHz $\varepsilon'$, we can conclude that $\mu' \approx 1$ at 300 GHz and no essential dielectric dispersion (i.e. no dielectric excitation) takes place in the MW range between 1 and 300 GHz, at least below 300 K. So, the frequency change of the electromagnetic parameters (Fig. 6b,c) is caused only by the MW magnetic dispersion. The high-frequency permeability dispersion was observed below 1 GHz at temperatures between 100 K and 400 K and related to the F1 and F2 modes, which were attributed to dynamics of the domain walls [28]. Above $T_{c2}$, the main high-frequency permeability dispersion takes place below 500 MHz, values of the F1 and F2 frequences are lower than 400 MHz, and temperature dependences of the permeability at 500 MHz, 1 GHz and 1.8 GHz are nearly identical and follow the same Curie-Weiss law. This means that the anomaly in the magnetic permeability seen at magnetic phase transitions must be due to excitations above 1.8 GHz. Therefore, we propose that the critical slowing down of the F4b mode is related to the Curie-Weiss behavior of the 1 GHz permeability. Of course, some magnetic dispersion could be present also above 1.8 GHz and other magnetic excitations could contribute to the 1 GHz permeability beside the F4 mode (most probably F7). Nevertheless, only the F4b mode critically depends on temperature near $T_{c2}$.

Note that $\varepsilon'\mu'(T)$ and $\varepsilon''\mu''/\varepsilon'\mu'(T)$ dependences at 5.8 GHz show sharp anomalies near $T_{c2}$. The $\varepsilon'\mu'(T)$ minimum (in fact, the $\mu'(T)$ minimum) could be explained by the critical slowing down of the mean F4b frequency (Fig. 5) below the frequency of our experiment, like that of the soft mode in ferroelectrics [40]. In the collinear phase above $T_{c2}$, at temperatures where F4b > 5.8 GHz, $\varepsilon'\mu'$ values at 1 and 5.8 GHz are nearly equal, so the MW magnetic dispersion takes place mainly above 5.8 GHz. In the conical phases below $T_{c2}$, the higher level of the 5.8 GHz parameters in comparison with the 300 GHz ones also evidences MW magnetic dispersion even above 5.8 GHz. It correlates with the presence of the magnetic excitations (magnons) in the THz range (see Fig. S7 in Supplementary material [36]).

We assume that all MW excitations in Fig. 5 are somehow related to the spin dynamics in CoZnU hexaferrites. MW excitations F4b and F7 with a very pronounced temperature dependence are well observed in both MSL and CPW (the A-modes are shown in Fig. 5 only for CPW to preserve the figure clarity). Above $T_{c2}$, the F7, F8 and F9 excitations join together into the F(7-9) one. The F6 excitation is well resolved only below and near $T_{c2}$, and absent above 315 K. It is characterized by the weak temperature dependence. The F5 excitation is better seen as a transmission mode below $T_{c2}$ and as an absorption mode above $T_{c2}$. It is characterized by a weak temperature dependence below $T_{c2}$, but its frequency (estimated from the $A(f)$ maxima) essentially increases with temperature above $T_{c2}$. The F3 excitation is well resolved only above 120 K, and its mean frequency is nearly temperature independent above 200 K at the 2 GHz level. Above 320 K, the F4a mode joins F3. The F3 frequency increases below 200 K towards $T_{c3}$. Let us note that the F3 mean frequency is close to the frequency of the F3 excitation observed at RT in the coaxial experiment [28], so we consider them as the same excitation.

The F4b mean frequency critically decreases towards $T_{c2}$. We compare its temperature dependence with that of MW inverse permeability at 1 GHz which obeys the Curie-Weiss law above $T_{c2}$:

$$1/\mu'(T) = 1/(\mu_L + \frac{C}{T-\theta}), \qquad (1)$$

with parameters: $\mu_L = 1.07$, $C = 8$ K and $\theta = 305$ K [28]. The temperature dependences of the F4b mean frequency from the CPW experiment (CPW-T in Fig. 5) and of the inverse 1 GHz permeability are shown together in Fig. 7. The F4b mean frequency is proportional to the inverse 1 GHz permeability with a constant factor of 25 GHz at temperatures above $T_{c2}$. It means that despite the permeability dispersion above 1 GHz, the $\mu'(T)$ follows the Curie-Weiss law up to ~22 GHz at high temperatures. Proportionality between the F4b frequency and inverse 1 GHz permeability (Fig. 7) proves the most important contribution of the F4b mode. For that reason, we can claim that the temperature dependence of the 1 GHz permeability above $T_{c2}$ is mainly caused by the temperature change of the F4b frequency. We can therefore assign the F4 (F4b) excitation to the critical magnetic excitation, which should correspond to the ferromagnetic resonance (FMR) in zero bias magnetic field (i. e. natural ferromagnetic resonance). Other two ferrimagnets with U-hexaferrite structure, $Ba_4Co_{2-3x}Cr_{2x}Fe_{36}O_{60}$ and $Ba_4Zn_{2-x}Co_xFe_{36}O_{60}$, have the FMR frequency at RT near 8 GHz, [26] and 9.4 GHz, respectively (the latter one in external magnetic field of 0.3 T) [41]. It supports our suggestion that the F4 mode in CoZnU corresponds to the FMR.

Based on the analysis described above, we can propose an explanation for the origin of the observed MW excitations. First, we consider all observed MW modes (F1 – F9) to be magnetic. The high number of the magnetic excitations is related to the complex magnetic structure of CoZnU consisting of (SRS*R*S*T)$_3$ magnetic blocks (note that CoZnU contains three formula units with a total of 108 Fe and 3 Co magnetic cations) [27]. It is responsible for the formation of many magnon branches in the Brillouin zone [44] and some of the magnons can be activated in MW and THz spectra. The F1, F2, F3, F4 (F4a, F4b), F7 and F8 modes are purely magnetic. Their mean frequencies correspond to the magnetic loss maxima or MW absorption maxima caused by interaction of the electromagnetic field with the magnetic excitations in the CoZnU hexaferrite. (No dielectric excitation takes place in the MW range between 1 and 300 GHz). The F1 and F2 modes are attributed to dynamics of the magnetic domain walls [27]. The origin of F3 is not so clear, we can attribute it to dynamics of the inhomogeneous magnetic structure of the non-magnetized hexaferrite ceramics with various orientations of the local magnetic moments of grains and domains, with the presence of grain boundaries.

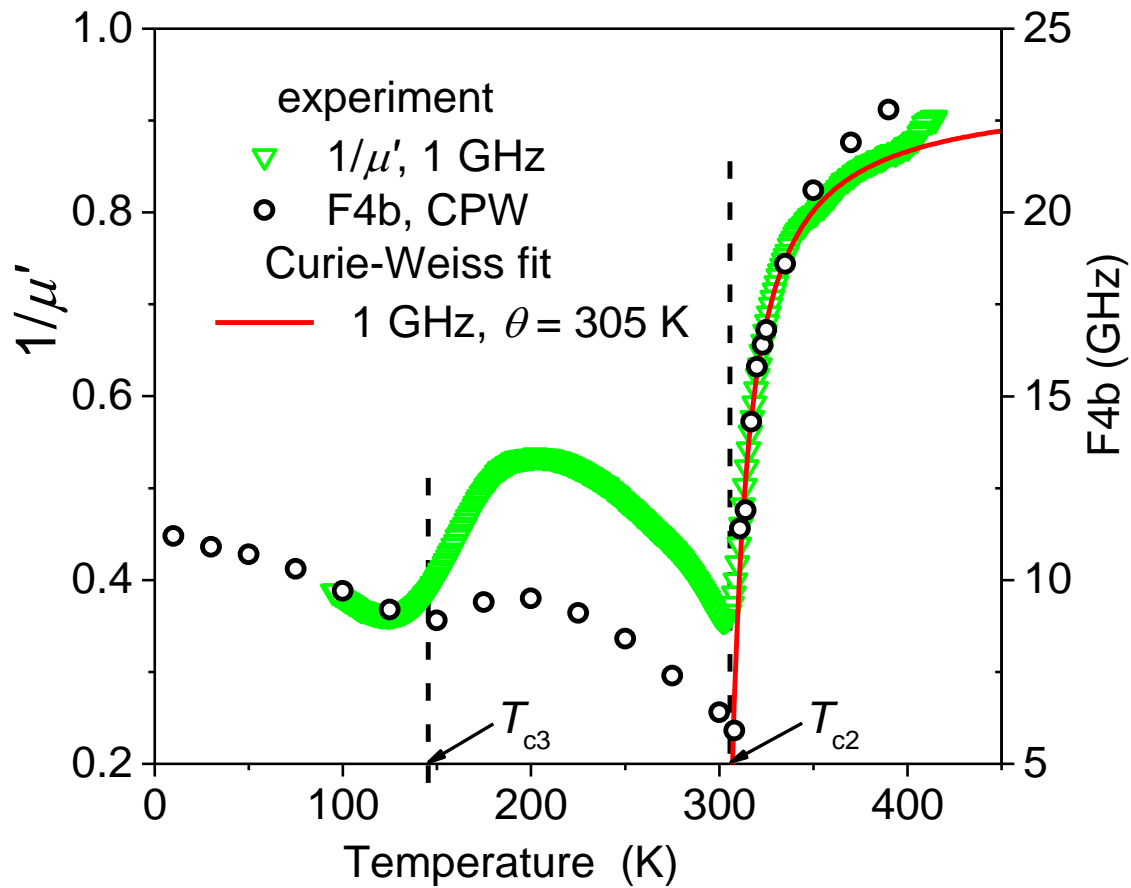

Fig. 7. Temperature dependences of the inverse 1 GHz magnetic permeability with its Curie-Weiss fit (Eq. 1) above 305 K and eigenfrequency of the F4b magnetic excitation corresponding to FMR.

The F4 mode – natural FMR (at zero magnetic bias) - is a spin wave (magnon) attributed to the collective spin dynamics. High frequency of the natural FMR in all ferrimagnetic phases evidences the existence of the non-zero internal magnetization and magnetocrystalline anisotropy in the previously not magnetized hexaferrite ceramics. The FMR frequency is unambiguously related to the magnetocrystalline anisotropy strength. Splitting of the natural FMR into two components (F4a and F4b) in the collinear phase is caused by a change of the magnetocrystalline anisotropy during the conical to collinear phase transition at $T_{c2}$.

The higher-frequency pure magnetic modes (F7 and F8) are other magnons of the exchange origin [42,43,44]. The F5, F6 and F9 modes are magnetodielectric with dominating influence of the magnetic properties on their temperature and field dependences. In the collinear phase, these modes disappear, join the FMR (F5, F6), or overlay with F7 and F8, creating the F(7 - 9) triplet.

**5. Magnetic nonlinearity of the microwave excitations.**

Essential influence of the weak magnetic bias field $H$ on the RT permeability dispersion, measured in the coaxial setup below 3 GHz, and the bias field dependences of the F1, F2 and F3 mean frequencies responsible for this dispersion, were reported in Ref. [28]. It motivated us to study the bias field influence on the MW magnetic excitations above 3 GHz, i.e., on the modes F3 – F9. The broadband spectra of the RT transmission of the CPW and MSL loaded with a CoZnU sample at different magnetic bias fields (Fig. 8) evidence the main evolution of the F3 – F9 modes with increasing $H$. The F4 $Tr(f)$ minimum near 6 GHz, corresponding to the FMR, changes non-monotonically. With increasing $H$, the minimum first becomes sharper, and its depth achieves a maximum of ~10 dB at ~300 Oe, then it broadens and its depth decreases (see also Fig. 9). The F4 $Tr(f)$ minimum is asymmetric at low $H$ and becomes symmetric at $H > 400$ Oe. It can be related to the presence of the F3 contribution at lower frequencies, which reduces with increasing $H$. The F4 frequency non-monotonically decreases with increasing $H$, with a local maximum at ~300 Oe corresponding to the deep minimum (Fig. 9).

The field $H > 400$ Oe essentially suppresses all MW modes above 8 GHz, i.e., F5 – F9. (see Fig. 8). The F5 – F9 modes in MSL are much more pronounced than in CPW (due to stronger coupling of the CoZnU sample with MSL), nevertheless, relatively weak $H = 400$ Oe reduces the depth of their minima down to the 4– 5 dB level, similar to that in CPW. Moreover, the peak system of the F7, F8 and F9 modes becomes more damped and finally these peaks merge above 300 Oe. Observed suppression of the F5 – F9 modes proves their magnetic origin. Their behavior under the magnetic bias reminds their temperature evolution on heating above the magnetic phase transition at $T_{c2} = 305$ K (see Fig. 5). Such a high sensitivity of all MW modes to the low bias field could be related to the low (hundreds of Oe) in-plane anisotropy field inducing rotation of the magnetization in the c-plane of hexaferrites [45].

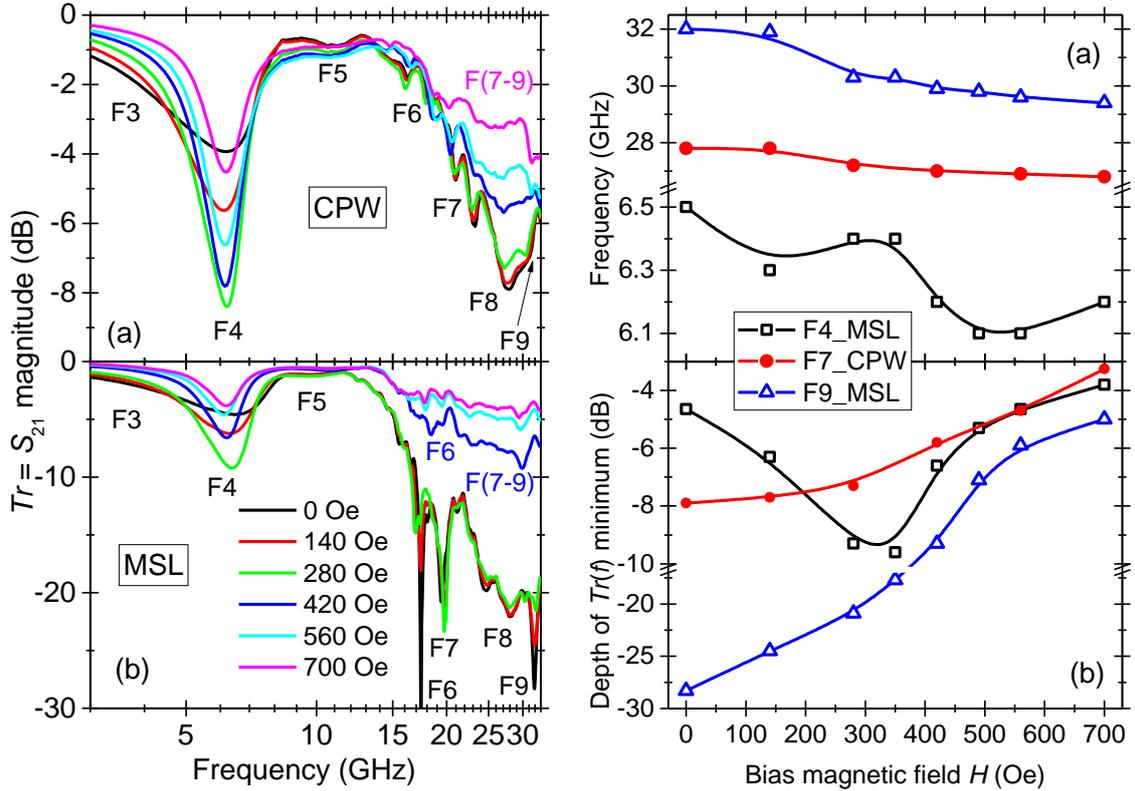

Fig. 8. RT transmission coefficient $Tr$ spectra of the CPW (a) and MSL (b) loaded with a CoZnU sample at different magnetic bias fields. The main $Tr(f)$ minima correspond to the MW excitations F3-F9. Note the log frequency scale.

Fig. 9. RT bias magnetic field dependences of frequencies (a) and depths (b) of the $Tr(f)$ minima corresponding to the MW excitations F4, F7 and F9 in CPW or MSL loaded with the CoZnU sample.

The RT reflection loss (*RL*) spectra of the CPW loaded with a CoZnU sample evidence the splitting of the FMR F4 mode into two components (F4a and F4b) under application of the weak bias $H$ field (Fig. 10), which is seen both in *RL* magnitude and phase. The 15 dB deep diffuse F4 $RL(f)$ minimum becomes sharp and 37 dB deep at $H = 140$ Oe, then splits into the 35 dB deep F4a and 28 dB deep F4a sharp minima at $H = 280$ Oe. At $H = 420$ Oe, the F4a minimum reduces to 20 dB, while the F4a minimum sharpens, and its depth increases to 35 dB. A subsequent increase of the bias field leads to the depth decreasing and widening of both minima. The evolution of the F4 frequency minimum and depth is shown in Fig. 11.

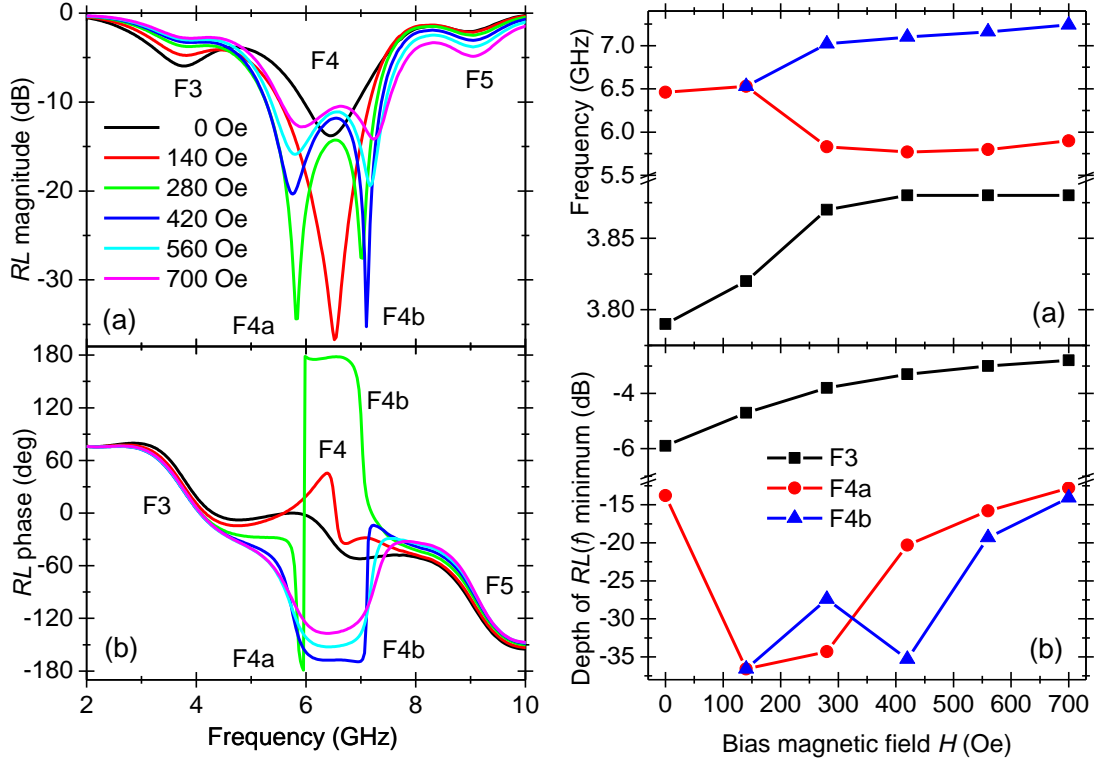

Fig. 10. RT reflection loss *RL* magnitude (a) and phase (b) spectra below 10 GHz of the CPW loaded with the CoZnU sample at different magnetic bias fields. The *RL* magnitude minima correspond to the MW modes F3, F4 and F5.

Fig. 11. RT bias magnetic field dependences of the frequencies (a) and depth (b) of the *RL* magnitude minima corresponding to the MW modes F4 and F3 in the CPW loaded with a CoZnU sample.

The *RL* spectra also confirm the presence of the F3 excitation near 4 GHz (Fig. 10) and its dependence on the applied bias field (Fig. 10). The F3 is seen also in the *Tr(f)* spectra (Fig. 8), but only as a contribution to the asymmetry of the diffuse F4 minimum. In the *RL* spectra, the F3 minimum is well separated from the F4 one. The application of the bias field suppresses F3 and reduces the depth of its *RL* minimum from 6 dB to 2.5 dB (Fig. 11b).

Let us note that the *RL* magnitude is used for the characterization of the electromagnetic MW absorption (*MA*) of materials [30,37,38]: *MA* = -*RL* (dB). So, the *RL(f)* minima correspond to the *MA(f)* maxima and, similar to the absorption coefficient *A(f)* maxima in the transmission experiment (Fig. 4), they are related to the loss maxima in studied materials. The comparison of the room-temperature *RL(f)* minima (Fig. 10a) at zero bias field and of 275 – 300 K *A(f)* maxima (Fig. 4b) allows to explain the very asymmetric shape of the *A(f)* maxima by the loss contributions from both F4 and F3 excitations. The loss dependence on the applied bias field proves the magnetic nature of the F3 and F4 excitations.

For quantitative analysis of the CoZnU magnetic nonlinearity, we use a simple method developed for the description of the giant magnetoimpedance effect [46], i.e. comparison of the *H*-dependent parameters ($Z_{in}$) with those under the maximum bias field $H_{max}$ applied in our experiment:

$$\frac{\Delta Z_{in}}{Z_{in}}(\%) = 100\% * \frac{[Z_{in}(H) - Z_{in}(H_{max})]}{Z_{in}(H_{max})} \tag{2}$$

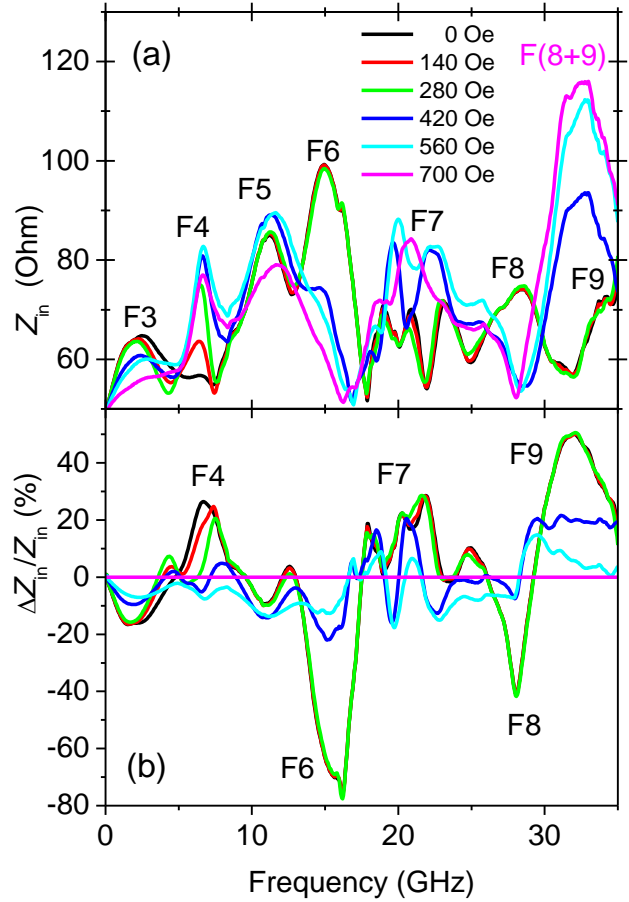

Fig. 12. Broadband spectra of the input impedance $Z_{in}$ of the CPW loaded with the CoZnU sample at different magnetic bias fields (a) and magnetoimpedance effect (b) measured at RT. The main $Z_{in}(f)$ maxima Z3– Z9 correspond to the MW excitations F3 – F9.

The broadband spectra of the input impedance $Z_{in}$ of the CPW loaded with the CoZnU sample are shown in Fig. 12a. They present remarkable dependence on the bias magnetic field in the whole studied frequency range, from 100 MHz to 35 GHz, especially near $Z_{in}(f)$ maxima corresponding to the MW modes F3-F9. The increase of the bias magnetic field suppresses the F3 and F6 maxima, nonmonotonically influences the F4 and F7 maxima, and results in increase and emergence of the F8 and F9 maxima. The magnetoimpedance effect (Fig. 12b) is the most pronounced near the F4 (> 20 %), F6 (> 70 %) and F8-F9 maxima or minima (± 50 %). Generally, the observed magnetoimpedance effect proves the magnetic nature of the main MW modes between 2 and 35 GHz.

In order to understand why the MW response of the CoZnU ceramics (transmission spectra, reflection loss spectra and input impedance) essentially changes with increasing temperature or under application of a weak magnetic bias field $H < 700$ Oe at RT, let us analyze magnetization (discussed in detail in [28]) in a weak field. The temperature dependence of the weak-field magnetization $M_W(T)$ at $H = 100$ Oe (Fig. 13a) shows a sharp and substantial anomaly at the phase transition from the conical to the collinear phase [28] at $T_{c2} = 305$ K, i.e., close to RT. The main changes of the magnetization $M(H)$ curves at RT and below it, i.e., in the conical magnetic phases, take place in a weak field $H < 700$ Oe and have similar step-like shape and value (Fig. 13b). This step corresponds to the weak-field magnetization $M_W$ [28]. The shape of the $M(H)$ curve at $H < 700$ Oe in the collinear phase, at 348 K, is not step-like; the $M_W(H)$ contribution is smaller. Such $M_W(H)$ temperature evolution corresponds to the $M_W(T)$ curve in Fig. 13a. The weak-field magnetization $M_W$ and its temperature or field evolution is attributed to the transformation from the polydomain to the monodomain magnetic structure [28]. At RT, i.e., near $T_{c2}$, the transformation from the polydomain conical structure to the polydomain collinear one with increasing weak bias field can also contribute to $M_W$. Beside $M_W$, the high-field $M_H$ contribution is seen at $H = 2\text{-}7$ kOe in the conical phase below $T_{c2} = 305$ K (Fig. 13b), which is attributed to the field-induced transition from the conical to the collinear magnetic phase [28].

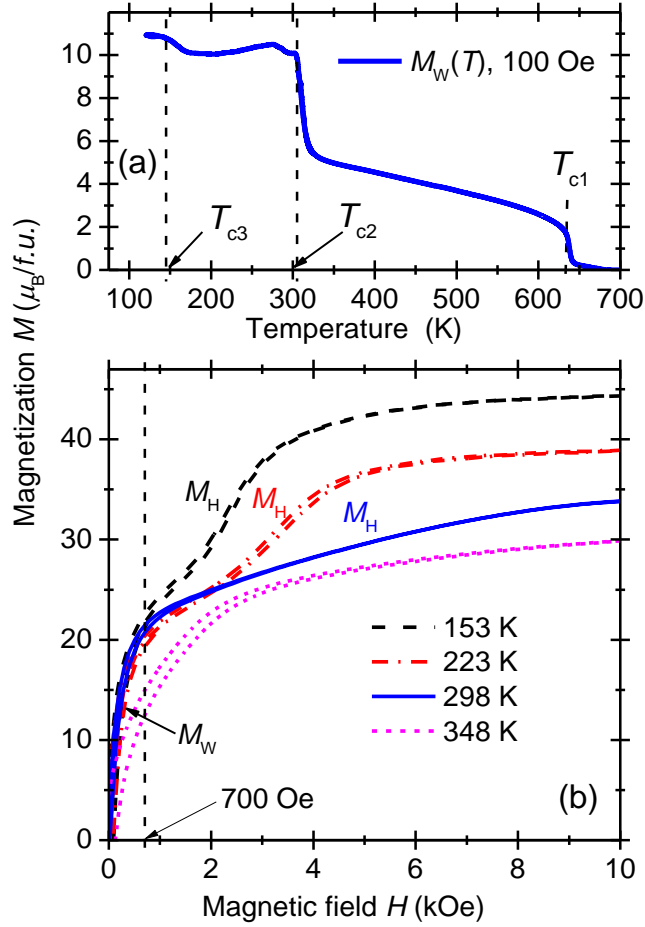

Fig. 13. Temperature dependence of the weak-field magnetization $M_w$ (a). Magnetic field dependences of the magnetization $M$ (b) at selected temperatures. (Data are adopted from [28].)

Overall, the highly nonlinear MW response and the high sensitivity of the main magnetic modes of the CoZnU ceramics to the weak magnetic bias field $H < 700$ Oe at RT, as well as the observed FMR mode (F4) splitting, are defined by the weak-field magnetization $M_w$ caused by the gradual transformation of the conical polydomain magnetic structure to the collinear polydomain magnetic one, and to the formation of the monodomain structure. It is not a result of the bias field induced phase transition which should take place at higher bias fields, above 1.5 kOe. The application of high bias magnetic field up to 10 kOe could reveal the $T$- and $H$-evolution of the MW excitation in the monodomain sample and during the monodomanization process. It will be a subject of our next study.

The high nonlinearity under the application of the low magnetic field makes the CoZnU hexaferrite ceramics interesting as a material for the highly tunnable MW filters [3,4].

## 9. Conclusions

We revealed 9 MW magnetic excitations below 35 GHz in the $Sr_4CoZnFe_{36}O_{60}$ hexaferrite ceramics and attributed them to dynamics of the magnetic domain walls, to the magnons and magnetodielectric modes. Anomalous behavior of the magnons was observed near the magnetic phase transitions. The lowest-frequency magnon was attributed to the natural FMR. Its high frequency (above 5 GHz in all ferrimagnetic phases) proves existence of the non-zero internal magnetization and magnetocrystalline anisotropy in the previously not magnetized ceramics. Splitting of the FMR into two components without magnetic bias was observed near the conical to collinear phase transition and attributed to a gradual transformation of the conical spin ordering to the collinear one during the phase transition at $T_{c2}$. The higher-frequency FMR component critically slows down to $T_{c2}$, especially on cooling in the collinear phase.

The FMR splitting was also revealed under the low bias (300 – 700 Oe) at room temperature (i.e. in the conical phase near $T_{c2}$), as well as the essential suppression of the higher-frequency magnons. The high sensitivity of the MW response to the weak magnetic bias field was shown to correlate with a weak-field magnetization of the CoZnU ceramics. This evidences that the FMR splitting under the low bias is caused by the gradual transformation of the conical polydomain magnetic structure to the collinear one and is not a result of the bias field induced phase transition which should take place at higher bias fields, at least above 1.5 kOe.

The high number of MW magnetic excitations sensitive to the weak magnetic field confirms the suitability of using our U-hexaferrite as MW absorbers, shielding materials and highly tunable magnetic materials. Our study confirms the reliability and effectivity of both used MW planar line techniques. We suggest that information on the genuine MW magnetic excitations in the U-hexaferrite could be useful for development of the MW elements based on the other types of the planar lines.

**Supplementary Material**

The supplementary material contains mainly additional supporting data from CPW, MSL, resonance and THz measurements.

**Acknowledgments**

This work has been supported by the Czech Science Foundation (Project No. 21-06802S), Grant Agency of the Czech Technical University in Prague (Project No. SGS22/182/OHK4/3T/14), by the Research Infrastructure NanoEnviCz (funded by MEYS CR, Projects No. LM2018124), and by project TERAFIT - CZ.02.01.01/00/22_008/0004594 co-financed by European Union and the Czech Ministry of Education, Youth and Sports.

**Declaration of Competing Interest**

The authors have no conflicts to disclose.

**Data availability**

The data that support the findings of this study are available from the corresponding author upon reasonable request.

**Author Contributions**

**M. Kempa:** Data curation (equal); Formal analysis (equal); Investigation (equal); Writing – review & editing (equal). **V. Bovtun:** Data curation (equal); Formal analysis (equal); Investigation (equal); Writing – original draft (lead); Writing – review & editing (equal). **D. Repček:** Investigation (equal); Writing – review & editing (equal). **J. Buršík:** Data curation (equal); Investigation (equal). **C. Kadlec:** Data curation(equal); Investigation (equal). **S. Kamba:** Writing – original draft (supporting); Writing – review & editing (equal); Conceptualization (lead); Funding acquisition (lead); Project administration (lead).

# Microwave magnetic excitations in U-type hexaferrite Sr$_4$CoZnFe$_{36}$O$_{60}$ ceramics
## *Supplemental material*

M. Kempa,[1] V. Bovtun,[1] D. Repček,[1,2] J. Buršík,[3] C. Kadlec,[1] S Kamba[1]

[1]Institute of Physics of the Czech Academy of Sciences, Na Slovance 2, Prague 182 00, Czech Republic

[2]Faculty of Nuclear Sciences and Physical Engineering, Czech Technical University in Prague, Břehová 7, 115 19 Prague, Czech Republic

[3]Institute of Inorganic Chemistry, Czech Academy of Sciences, 250 68 Řež near Prague, Czech Republic

**S1. Broadband microwave transmission and absorption.**

The full set of *S*-parameters ($S_{11}$, $S_{22}$, $S_{21}$, $S_{12}$, both magnitude and phase) of the microstrip line (MSL) and coplanar waveguide (CPW) loaded with CoZnU ceramics recorded in the broad frequency range 10 MHz - 50 GHz from 10 K to 390 K shows a reciprocal agreement of the $S_{11}$ and $S_{22}$, $S_{21}$ and $S_{12}$ spectra. It proves the high quality of the calibration procedure and the reliability of the experiment. For further analysis, the $S_{21}$ and $S_{11}$ spectra were mainly used. As the most sensitive and informative, the following parameters were selected and calculated [1,2,3]:

a) $S_{21}$ magnitude in dB (transmission coefficient *Tr* corresponding to the shielding efficiency *SE* = - *Tr*);

b) input impedance $Z_{in}$ of the MSL or CPW loaded with CoZnU samples, which can be calculated from the $S_{11}$ magnitude in dB (reflection coefficient *R*) as

$$Z_{in} = Z_0 \frac{1+10^{S_{11}/20}}{1-10^{S_{11}/20}}, \qquad (s1)$$

where $Z_0$ = 50 Ohm is an input impedance of the unloaded, empty MSL or CPW;

c) absorption coefficient *A* which can be calculated as

$$A = 20 \log \left(1 - \left|\frac{2Z_{in}}{Z_{in}+Z_0}\right| - \left|\frac{Z_{in}-Z_0}{Z_{in}+Z_0}\right|\right), \tag{s2}$$

The experimentally recorded $S_{11}$ ($S_{11}$ magnitude in dB is a reflection coefficient $R$) and $S_{21}$ ($S_{21}$ magnitude in dB is a transmission coefficient $Tr$) spectra of the CPW loaded with a CoZnU ceramics (see Fig. S1) show multiple resonances with the phase change between –180 and +180 degrees, $R(f)$ maxima above -10 dB and $Tr(f)$ minima below -10÷20 dB. The unloaded CPW is characterized (after the calibration) by the parameter values $R < $ -30 dB, $Tr \approx 0$ dB and $S_{21}$ phase $\approx 0$ degrees in the whole frequency range of the experiment. So, all observed resonances are caused by the CoZnU sample coupled with the CPW transmission line. The change and the frequency shift of the observed resonances with temperature are much larger than those caused by the temperature change of the sample and the line sizes, and should be related to the temperature evolution of the sample's electromagnetic properties. The most temperature dependent $S_{21}$ resonance (transmission mode) is marked as T4 in Fig. S1c. $S$-parameters spectra of the MSL loaded with a CoZnU sample slightly differ from those of CPW (due to the different electromagnetic field distribution and sample coupling) but show qualitatively similar temperature changes. This similarity also supports the attribution of the $S$-parameters temperature change to the temperature evolution of the CoZnU electromagnetic properties. $S$-parameters spectra are noisy above 35 GHz, therefore we evaluate data mainly below 35 GHz (see Fig. 1 – Fig. 4 of the main text).

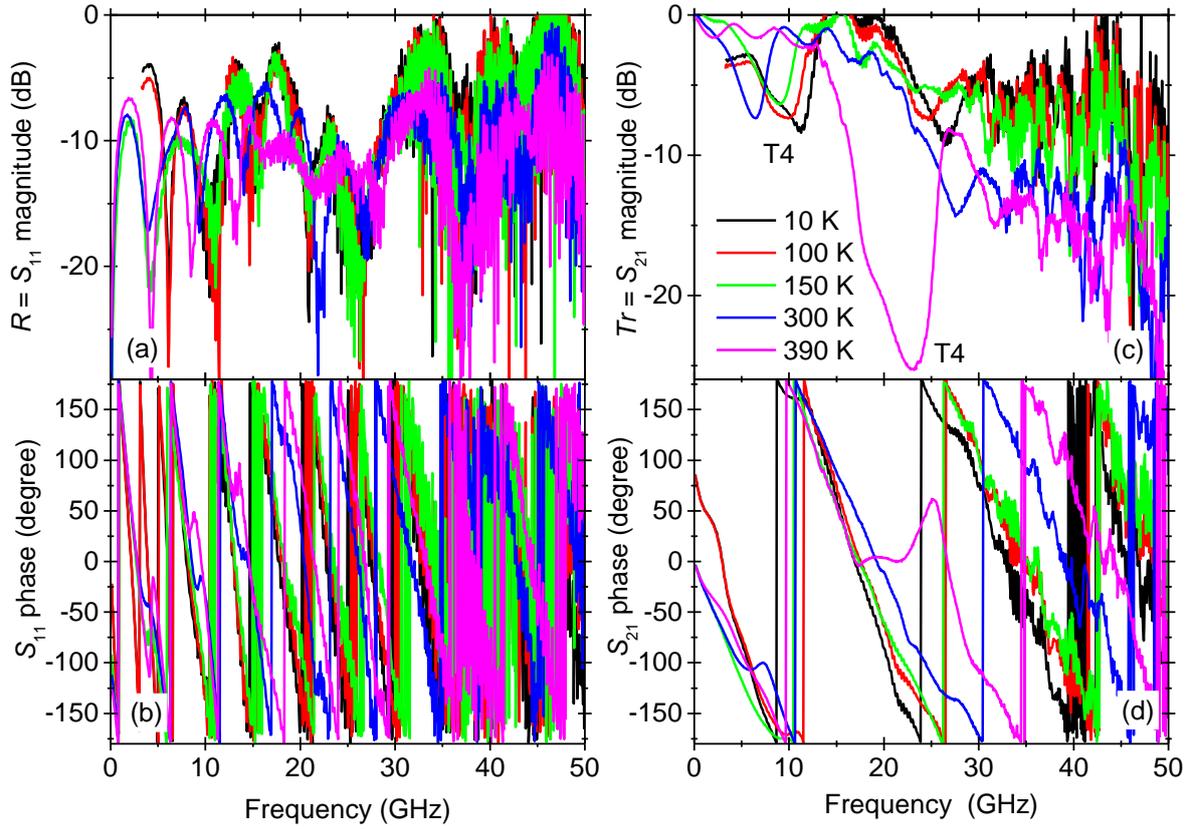

Fig. S1. Recorded broadband spectra of the $S_{11}$ (a, b) and $S_{21}$ (c, d) parameters of the CPW loaded with a CoZnU sample at selected temperatures.

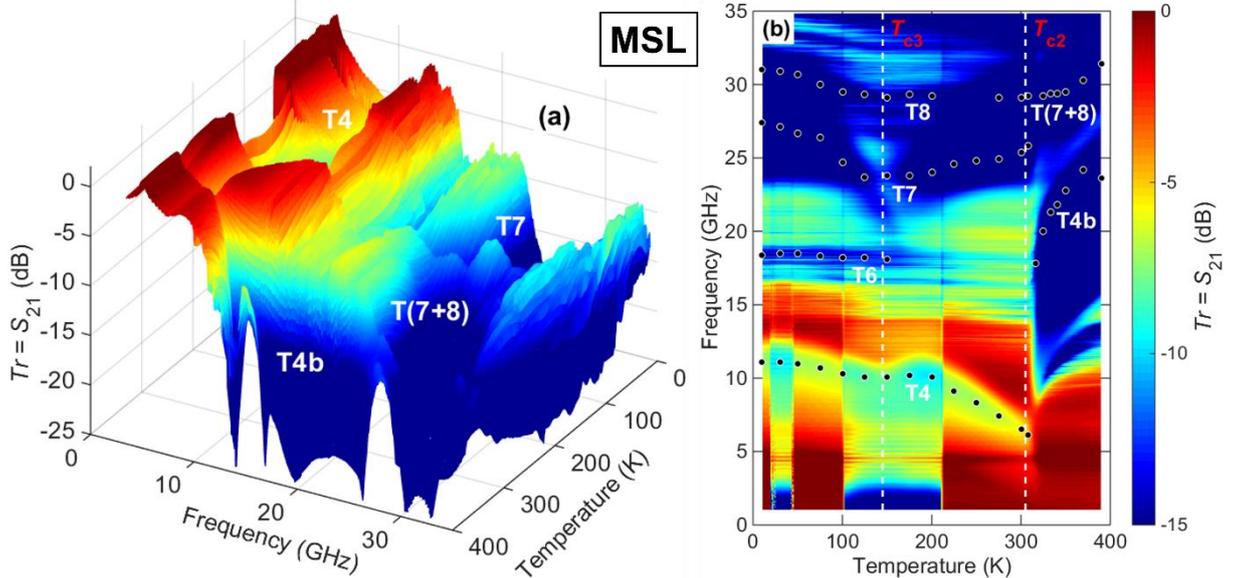

The same set of the transmission modes is observed in both CPW and MSL experiments, with the same CoZnU sample. Moreover, the temperature evolution of the CPW (Fig. 3 in the main text) and MSL (Fig. S3) T-modes is similar, with a remarkable change of their behavior near the CoZnU magnetic phase transitions.

Fig. S2. Temperature-frequency maps of the transmission coefficient $Tr$ ($S_{21}$ magnitude) of CoZnU ceramics obtained using the MSL: (a) in 3D view, (b) as a pseudo-3D color map. In (b), points correspond to the main $Tr(f)$ minima shown in Fig. S3. Temperatures of the phase transitions are marked by dashed lines. Apparent jumps of the $Tr$ at several temperatures are instrumental artifacts.

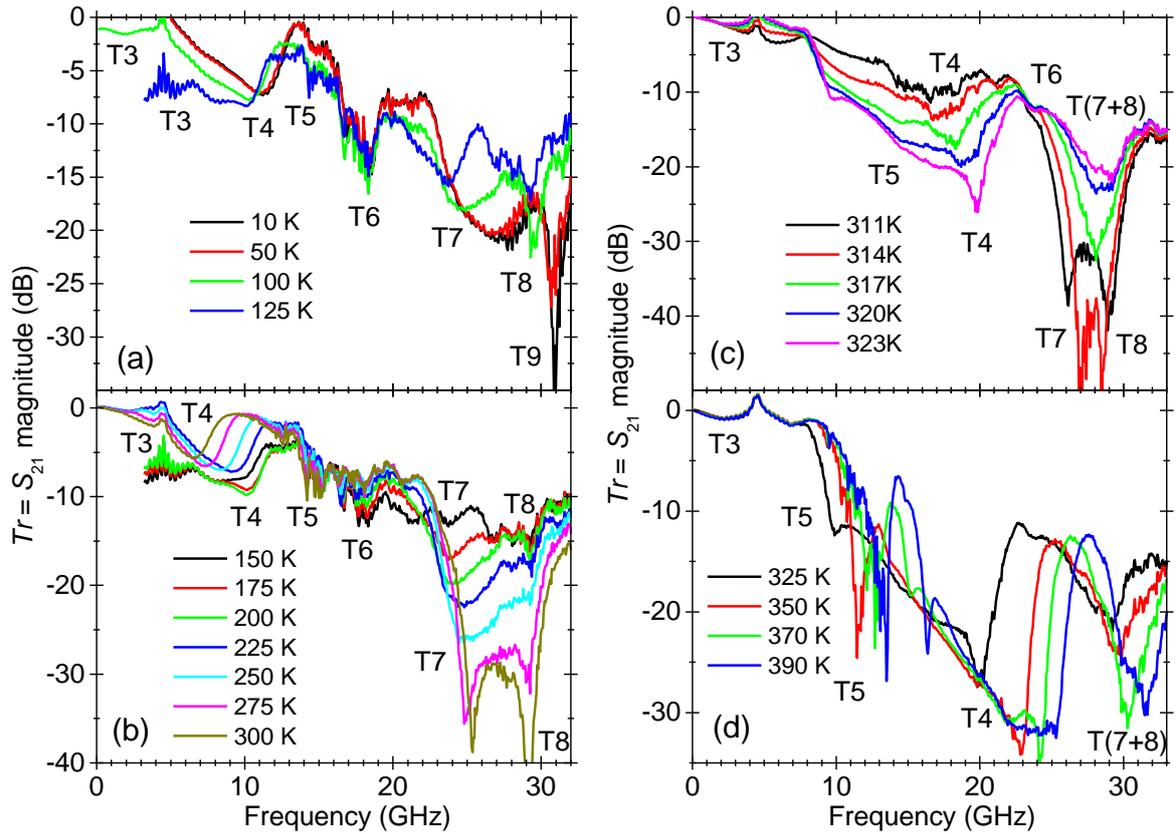

Fig. S3. Broadband spectra of the transmission coefficient $Tr$ ($S_{21}$ magnitude) of the MSL loaded with a CoZnU sample at selected temperatures below $T_{c3}$ (a), between $T_{c3}$ and $T_{c2}$ (b), and above $T_{c2}$ temperature (c, d). The main $Tr(f)$ minima are marked T3 - T9.

In the MSL absorption spectra (Fig. S4), 6 diffuse $A(f)$ maxima can be recognized below 35 GHz, and they are numbered from A4 to A9 with increasing frequency. These maxima evidence microwave excitations (absorption modes, A-modes). The same system of the A-modes and their similar temperature evolution was derived from both CPW and MSL experiments (except for the A3 mode, which is weak and absent in the MSL spectra).

Temperature behavior of the A-modes remarkably changes near the magnetic phase transitions, similar to those of T-modes.

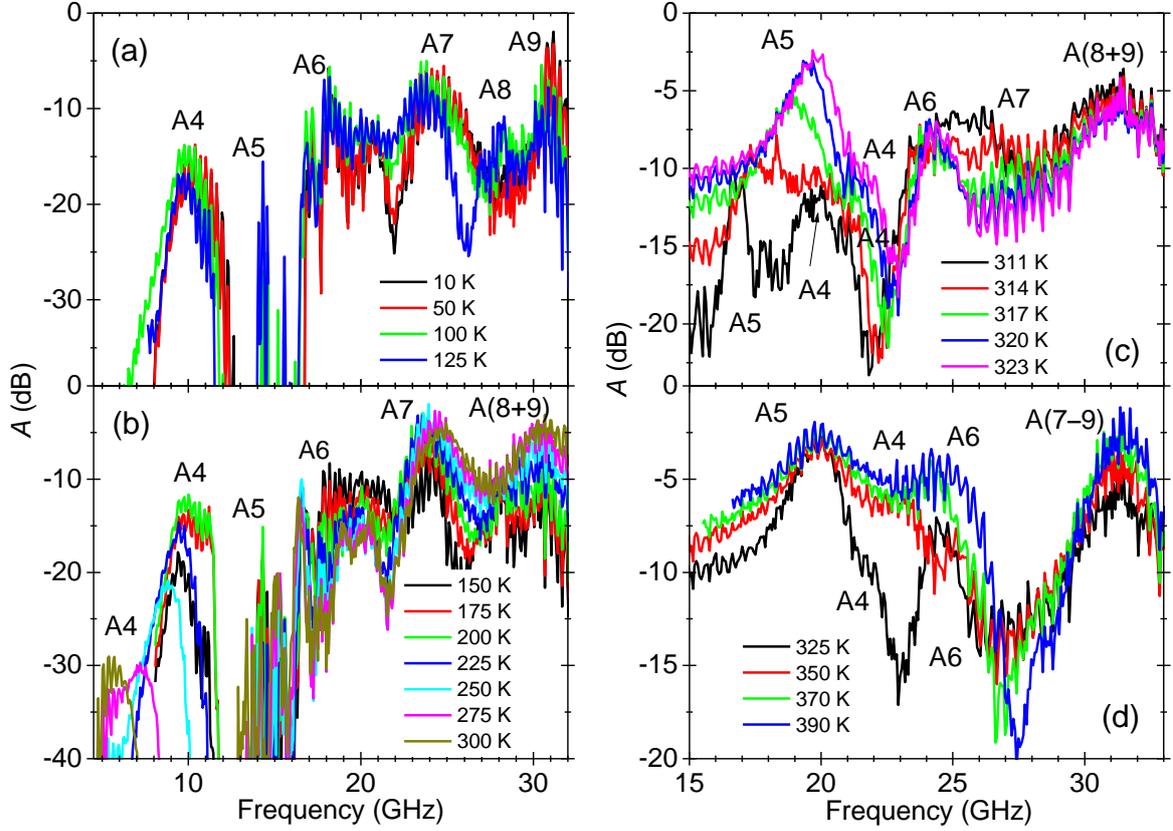

Fig. S4. Broadband spectra of the absorption coefficient $A$ of the MSL loaded with a CoZnU sample at selected temperatures below $T_{c3}$ (a), between $T_{c3}$ and $T_{c2}$ (b), and above $T_{c2}$ (c, d). The main $A(f)$ maxima are marked A4 - A9.

We also analyzed the input impedance spectra $Z_{in}(f)$, calculated using Eq. s1 and shown for CPW in Fig. S5. The spectra for MSL look very similar. The impedance spectra are noisy, similar to the absorption ones (Fig. S4), but less diffused. 7 main $Z_{in}(f)$ maxima seen below 35 GHz are marked Z3 - Z9. These maxima correspond to the transmission minima T3÷T9 (Fig. 2 in the main paper). Sometimes, the $Z_{in}(f)$ maxima are more pronounced, less diffuse, and better separated than the $Tr(f)$ minima. The impedance spectra also confirm a remarkable change of the T-modes behavior near the magnetic phase transition at $T_{c2}$ in the CoZnU ceramics.

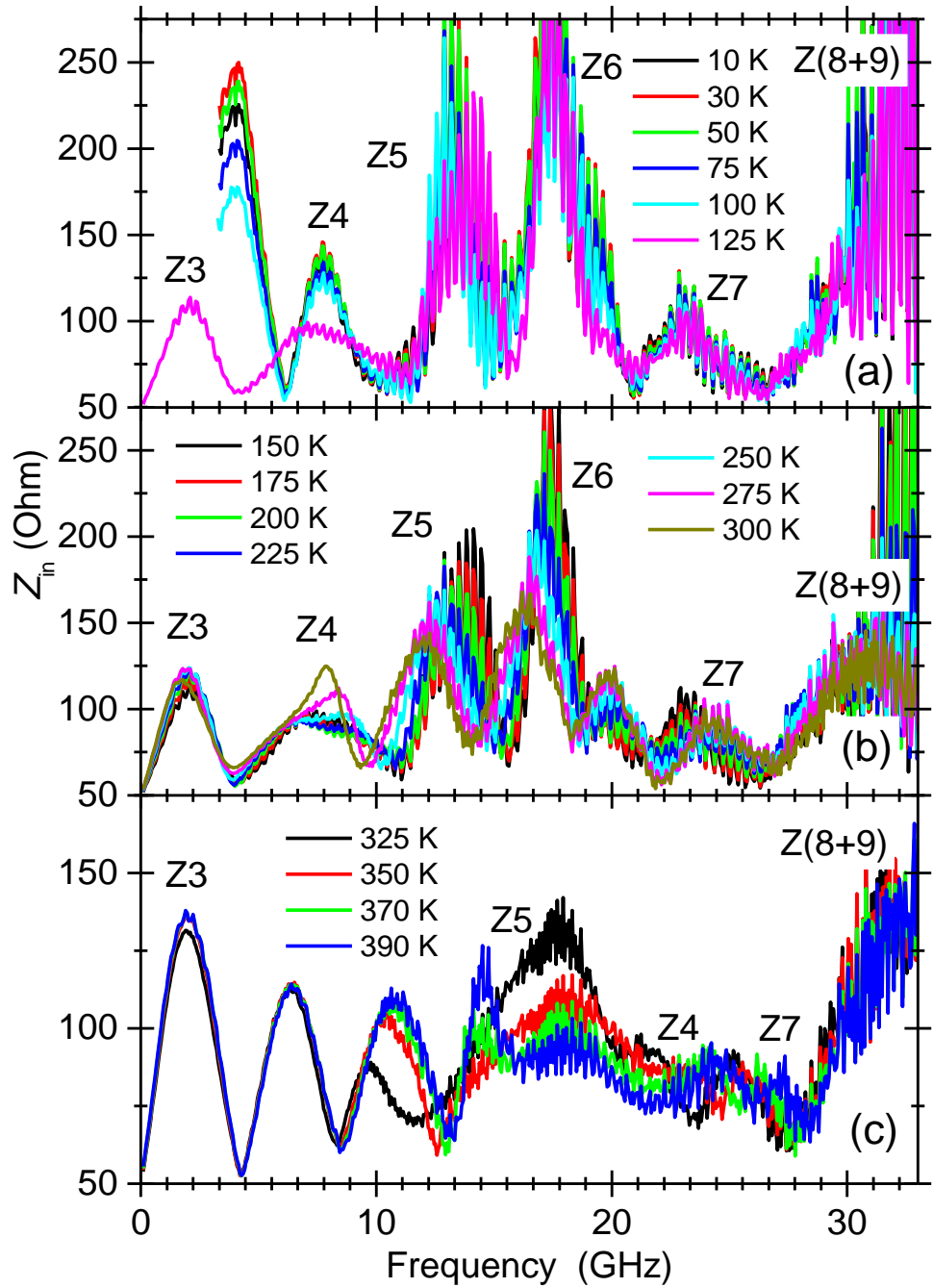

Fig. S5. Broadband spectra of input impedance $Z_{in}$ of the CPW loaded with a CoZnU sample at selected temperatures below $T_{c3}$ (a), between $T_{c3}$ and $T_{c2}$ (b), and above $T_{c2}$ (c). The main $Z_{in}(f)$ maxima are marked Z3 - Z7.

The transmission spectra of the thin cylindrical CoZnU sample, measured as a dielectric resonator (DR) in the shielding cavity [4,5], show a few temperature-dependent resonances ($S_{21}$ maxima) between 8 and 16 GHz with an anomalous behavior near $T_{c2} = 305$ K (see temperature-frequency maps in Fig. S6). Opposite to the transmission spectra of CPW or MSL, the DR resonance frequencies (DR-T modes) correspond to the $S_{21}$ maxima, not minima, because the effective transmission is possible only in the case of sufficient coupling at the resonance frequency. The DR-T modes with the most pronounced $S_{21}$ maxima ($\geq$ -40 dB, dark red to brown colors) at ~10 GHz, ~11 GHz, ~13.5 GHz and ~15.5 GHz, and with weak $S_{21}$ maxima ($\geq$ -60 dB, green to yellow colors) are shown also as black points in Fig. S6. Of course, the mean frequencies of the CPW-T and DR-T modes are different, but their temperature behavior is similar. All DR-T modes below 14 GHz slow down toward $T_{c2}$, similarly to the CPW and MSL T-modes below 22 GHz (Fig. 4 in the main paper). The F4 excitation (CPW-T4) mode, which is characterized by the high absorption and the strongest temperature dependence above $T_{c2}$, is also shown as a solid line in Fig. S6. It crosses all DR-T modes near $T_{c2}$ with increasing temperature, and the crossing points correspond well to the DR-T mode temperature anomalies.

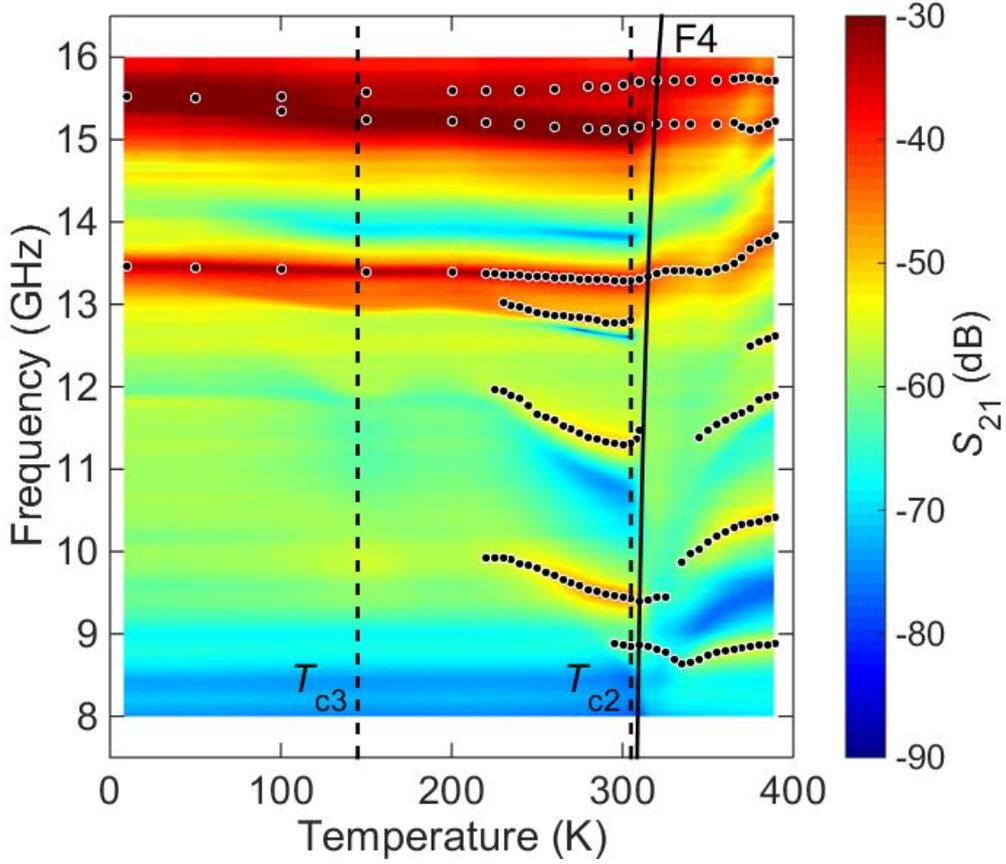

Fig. S6. Temperature-frequency map of the transmission coefficient ($S_{21}$ magnitude) of the CoZnU dielectric resonator. Points correspond to the main $S_{21}(f)$ maxima, solid line near $T_{c2}$ corresponds to the softening of the F4 frequency towards $T_{c2}$ (CPW-T curve from Fig. 4 in the main paper).

**S2. THz spectra.**

The time-domain THz transmission spectra were measured down to 4 K. The amplitude and phase of the passing wave were determined, and this allowed the complex refractive index $N = (\varepsilon^*\mu^*)^{1/2}$ to be calculated directly [6]. Seven excitations are visible in the THz spectrum, but since we are not sure which are magnetic and which are polar (phonon) excitations, we present in Fig. S7 only the components of the complex refractive index $N$ and not the complex dielectric permittivity $\varepsilon^*$ and magnetic permeability $\mu^*$ spectra. However, we estimate from the shape of the excitations which ones are magnons and which are phonons. We label their frequencies in Fig. S7 with arrows.

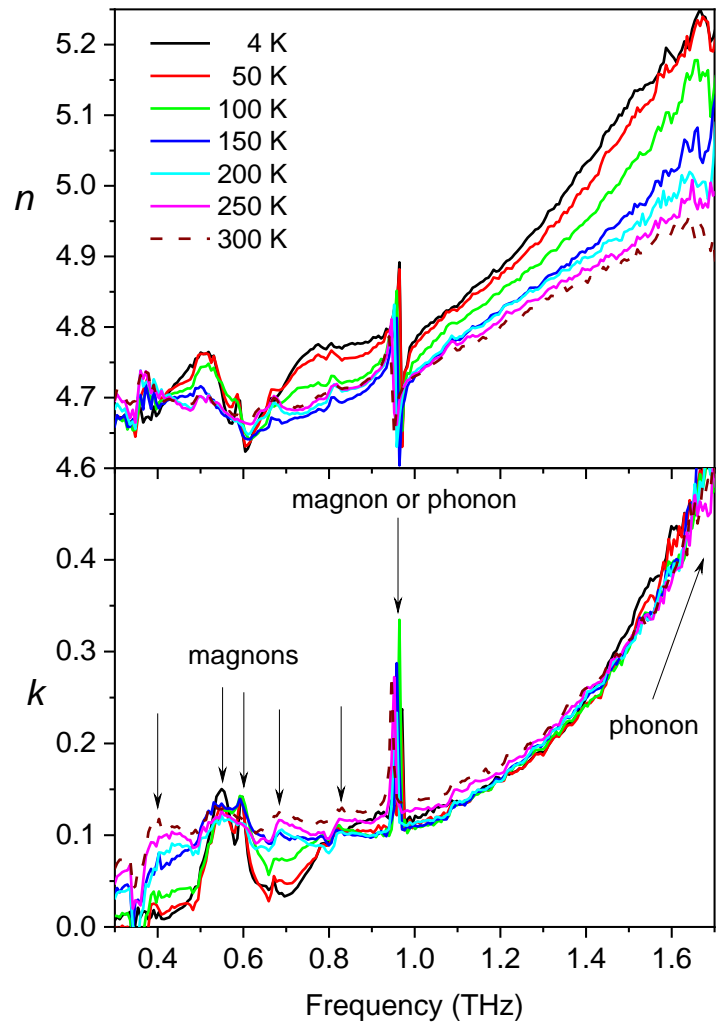

Fig. S7. Temperature dependence of the complex refractive index $N=n+\mathrm{i}k$ in the THz region. The frequencies of the magnetic and of the lattice excitations are marked.

**References.**